\documentclass[manuscript]{aastex6}

\usepackage{CJK}
\usepackage{amsmath}

\slugcomment{Not to appear in Nonlearned J., 45.}

\shorttitle{PS1 Neptune Trojans}
\shortauthors{Lin et al.}

\begin{document}
\begin{CJK*}{UTF8}{bkai}

\title{The Pan-STARRS 1 Discoveries of five new Neptune Trojans}

\author{Hsing Wen Lin (林省文)\altaffilmark{1}}
\affil{Institute of Astronomy, National Central University, 32001, Taiwan}
\email{edlin@gm.astro.ncu.edu.tw}

\author{Ying-Tung Chen (陳英同)\altaffilmark{2}}
\affil{Institute of Astronomy and Astrophysics, Academia Sinica, P. O. Box 23-141, Taipei 106, Taiwan}

\author{Matthew J. Holman\altaffilmark{3}}
\affil{Harvard-Smithsonian Center for Astrophysics, 60 Garden Street, Cambridge, MA 02138, USA}

\author{Wing-Huen Ip (葉永烜)\altaffilmark{1,4}}
\affil{Institute of Astronomy, National Central University, 32001, Taiwan}

\author{M. J. Payne\altaffilmark{3},  P. Lacerda\altaffilmark{6}, W. C. Fraser\altaffilmark{6}, D. W. Gerdes\altaffilmark{5}, A. Bieryla\altaffilmark{3}, Z.-F. Sie (謝宗富)\altaffilmark{1}, W.-P. Chen (陳文屏)\altaffilmark{1}\\ and\\W. S. Burgett\altaffilmark{7}, L. Denneau\altaffilmark{7}, R. Jedicke\altaffilmark{7}, N. Kaiser\altaffilmark{7}, E. A. Magnier\altaffilmark{7}, J. L. Tonry\altaffilmark{7},  R. J. Wainscoat\altaffilmark{7}, C. Waters\altaffilmark{7}}

\altaffiltext{1}{Institute of Astronomy, National Central University, 32001, Taiwan}
\altaffiltext{2}{Institute of Astronomy and Astrophysics, Academia Sinica, P. O. Box 23-141, Taipei 106, Taiwan}
\altaffiltext{3}{Center for Astrophysics, 60 Garden Street, Cambridge, MA 02138, USA}
\altaffiltext{4}{Space Science Institute, Macau University of Science and Technology, Macau}
\altaffiltext{5}{Department of Physics, University of Michigan, Ann Arbor, MI 48109, USA}
\altaffiltext{6}{Astrophysics Research Centre, Queen's University Belfast, BT7 1NN, Northern Ireland, UK}
\altaffiltext{7}{Institute for Astronomy, University of Hawaii at Manoa, Honolulu, HI 96822, USA}

\begin{abstract}
In this work we report the detection of seven Neptune Trojans (NTs) in the Pan-STARRS 1 (PS1) survey. Five of these are new discoveries, consisting of four L4 Trojans and one L5 Trojan. Our orbital simulations show that the L5 Trojan stably librates for only several million years. This suggests that the L5 Trojan must be of recent capture origin. On the other hand, all four new L4 Trojans stably occupy the 1:1 resonance with Neptune for more than 1\,Gyr. They can, therefore, be of primordial origin. Our survey simulation results show that the inclination width of the Neptune Trojan population should be between $7^{\circ}$ and  $27^{\circ}$ at $>$ 95\% confidence, and most likely $\sim 11^{\circ}$. In this paper, we describe the PS1 survey, the Outer Solar System pipeline, the confirming observations, and the orbital/physical properties of the new Neptune Trojans.

\end{abstract}


\section{Introduction}

The best known Trojans are the asteroids in a co-orbital 1:1 mean motion resonance with Jupiter. Those in stable libration around the Lagrange point $60^\circ$ ahead of Jupiter are called L4 Trojans, and those around the Lagrange point $60^\circ$ behind are called L5 Trojans. There are more than 6000 known Jovian Trojans with sizes $\gtrsim$ 10\,km. \citet{yos05} estimated that the total number of 1\,km sized Jovian Trojan could be as many as 600,000. After Jupiter, Neptune has the second largest population of Trojans. Prior to this study, nine L4 Neptune Trojans (or NTs) and three L5 NTs had been discovered \citep{ale14, ger16, par13, she06, she10a, ell05}.

\citet{nes02} examined the orbital evolution and long-term stability of Trojans of Saturn, Uranus, and Neptune, under the current planetary configuration. They found that unlike the cases of Saturn and Uranus, where their Trojans could be removed on relatively short time scales, the primordial population of NTs can survive to the present time after their formation. Subsequently, the first Neptune Trojan, 2001 QR$_{322}$ at L4, was found in the Deep Ecliptic Survey \citep{ell05}. Based on the low inclination  ($\sim 1.3^\circ$) of 2001 QR$_{322}$ with a size of $\sim$100 km, \citet{chi05} proposed that large ($\sim$100 km sized) NTs might be primordial objects formed in-situ by accretion in a thin disk. This means that NTs should generally have {\it i} $\lesssim 10^\circ$. Following the discovery of three more NTs, one of which has a high inclination (see Table~\ref{tab1} with a list of the known NTs and those detected in this study), \citet{she06} suggested that a thick cloud of high-inclination NTs which could be of capture origin, should exist with a 4:1 ratio over the low-inclination population.  

In the context of the Nice model \citep{gom05, tsi05}, \citet{mor05} investigated the chaotic capture of small bodies at the two Lagrangian points of Jupiter during the planetary migration phase. Following a similar approach, \citet{nes09} produced a model calculation of the capture process of NTs. Although the inclinations of the objects captured from the thin solar nebula disc could be later increased by dynamical processes, the numerical results could not account for the 4:1 high-{\it i} to low-{\it i} NTs ratio indicated by the observations of \citet{she06}. This discrepancy might be worsened if the orbits of the planetesimals before chaotic capture were excited by the gravitational scattering effect of a population of Pluto-sized objects according to these authors.    

\citet{par15} applied a statistical method to debias the observed distributions of orbital inclinations, eccentricities and libration amplitudes of NTs. His treatment confirmed the existence of the thick cloud population with $\sigma _i > 11^\circ$.  Here, $\sigma _i$ is the inclination width of the Brown's distribution \citet{bro01}:
\begin{equation}
p(i) = \sin(i)\exp(-\frac{1}{2}(i/\sigma_{i})^2)di
\end{equation}

From a numerical study of the resonant capture effect via planetary orbital migration, \citet{par15} showed that low-inclination objects can be captured into high-inclination NTs, but the conversion efficiency is too low to account for the presence of the high-inclination population. On the other hand, if the original planetesimals were characterized by high-inclination orbits, their NT-counterparts captured into 1:1 resonance with Neptune could preserve their high inclinations, and hence cause the formation of a thick NT cloud. 

\citet{che16} provided an alternative mechanism to effectively form the high-inclination NTs. They investigated how planetary migration affects the orbital elements distribution of NTs, and found that if orbital eccentricities and inclinations of Neptune and Uranus were damping during planetary migration, the secular resonances with Neptune will increase the probability of trapping the test particles into high inclination NT orbits. Moreover, most of primordial NTs, especially the high inclination ones, were unstable and lost in the damping case. From these results, their concluded that the current existent NTs can be explained by the capture origin, particular the trapping scenario with orbital damping of Neptune and Uranus during planet migration.

The first Neptune Trojan at L5,  2008 LC$_{18}$, was discovered by \citet{she10a}. One more was found by \citet{par13}. According to \citet{she06} and \citet{par15}, the difference in the numbers of known NTs in the L4 and L5 points, respectively, could be an observational bias caused by the fact that the L5 point of the NTs is currently in the vicinity of the Galactic center, making it difficult to clearly identify slowly moving foreground objects. 

Due to the small number of known NTs, it has been difficult to reconstruct their size distribution and to estimate their total number. \citet{chi05} and \citet{she06} suggested that the number of large (size $>$ 65 km) NTs should exceed that of the Jovian Trojans by more than a factor of ten. \citet{ale14} discovered one temporary and one stable NTs and derived the populations of $210^{+900}_{-200}$ and $150^{+600}_{-140}$, respectively, with $H \lesssim 10.0$.
From an ultra-deep, pencil-beam survey with a detection efficiency of 50$\%$ for objects with  R $\sim 25.7$ mag, \citet{she10b} derived that the cumulative luminosity function of $m_R < 23.5$ mag follows a steep power law of index $\alpha \sim 0.8 \pm 0.2$:
\begin{equation}
\Sigma(m_R) = 10^{0.8(m_R-m_0)}.
\end{equation}
In other words the size frequency distribution of the bright NTs at size a $>$ 100 km have a power-law index $\sim 5 \pm 1$:
\begin{equation}
dN/da \propto a^{-5}.
\end{equation}
For reference, Jovian Trojan population, cold population and hot population of TNOs have $\alpha \sim$ 1.0, 1.5 and 0.87, respectively \citep{fra14}. 
It clearly shows that the luminosity function of NT population has power law index, $\alpha$, similar to the Jovian Trojans and hot population of TNOs. This result is obvious interpretation from the fact that they all have the same size frequency distribution.

The long-term orbital stability of NTs has been studied by \citet{nes02}, \citet{dvo07} and \citet{zho09, zho11} who showed that NTs can be stable for over 4 Gyr even with orbital inclinations $\sim 30^\circ$. However, the stable region is  restricted in eccentricity ($e\lesssim 0.1$). The orbital stability of individual known NTs has been investigated by \citet{bra04}, \citet{gua12}, \citet{mar03}, \citet{hor10a}, \citet{hor12a}, \citet{hor12b} and \citet{lyk09}.  In general, they can be classified into three different dynamical regimes: 
\begin{enumerate}
\item Objects temporarily captured into unstable orbits: these kind of NTs are located completely outside the stable region and have a dynamical lifetime as short as 1 Myr \citep{hor12b, gua12}.
\item Objects in marginally stable orbits: these NTs are found near the edge of the stable region or in the proximity of the secular resonances with a dynamical life time of about 100 Myr \citep{hor10a, lyk11, zho11}.
\item Stable objects: they are located deep inside the stable region with a dynamical life time as long as the age of the Solar System and could be of primordial origin. 
\end{enumerate}

Our current knowledge of the NTs is based on the discoveries by several different surveys \citep{ale14, ell05, ger16, par13, she10a, she06}. Without a comprehensive full-sky survey to cover most of the Trojan clouds, it is difficult to estimate the total number, the size distribution, the orbital distribution and the L4/L5 asymmetry of Neptune Trojans. In comparison, the PS1 project covering the whole Northern Hemisphere to a limiting magnitude of $r_{\rm P1}\sim 22$ presents an ideal opportunity to search for NTs with significant reduction in the latitudinal and longitudinal biases. In this paper, we report the detections of seven NTs by PS1, five of which are new discoveries. 

This paper is organized as follows. Section 2 will introduce the PS1 survey and the Outer Solar System pipeline for searching of distant moving objects. In Section 3, we describe how to select, confirm the Trojan candidates and report the discoveries. In Section 4, we calculate the orbital and physical properties of the NTs. In Section 5, we describe how to perform the inclination debiasing of PS1 survey and investigate the intrinsic inclination distribution of stable L4 NTs. In Section 6, we roughly estimate the luminosity function of stable L4 NTs. In Section 7, we discuss the ratio of high- and low-inclination populations of NTs, and the possible asymmetry of L4 and L5 distributions. A summary is given in Section 8.

\section{Pan-STARRS 1 survey and the Outer Solar System pipeline}

The PS1 Survey began in May 2010 and ended in May 2014. 
With a 1.8-m Ritchey-Chretien reflector located on Haleakala, Maui, and a 1.4 gigapixel camera covering 7 square degrees on the sky, the PS1 telescope was able to observe the whole visible sky within a week, searching for all kinds of astrophysical transients and Solar System moving objects. 

The PS1 observations were taken using five different survey modes \citep{kai10}: 
\begin{enumerate}
\item The $3\pi$ Steradians Survey using PS1 photometric system \citep{ton12}, $g_{\rm P1}$ (bandpass $\sim$ 400-550 nm), $r_{\rm P1}$ ($\sim$ 550-700 nm), $i_{\rm P1}$ ($\sim$ 690-820 nm), $z_{\rm P1}$ ($\sim$ 820-920 nm) and $y_{\rm P1}$ ($\sim$ 920-1100 nm), which is similar but not identical to the SDSS/Sloan system with the addition of $y_{\rm P1}$. 
\item The Solar System Survey which is optimized for Near-Earth asteroids and other Solar System objects by covering the whole  $\pm 10^\circ$ and part of $\pm 20^\circ$ areas of the ecliptic plane with the $w_{\rm P1}$-band filter (400-820 nm) which is equivalent to $g_{\rm P1}+r_{\rm P1}+i_{\rm P1}$ . 
\item The Medium Deep Survey covering 10 selected fields and nightly observations with long exposures (113 sec for $g_{\rm P1}$ and $r_{\rm P1}$, 240sec for $i_{\rm P1}$, $z_{\rm P1}$ and $y_{\rm P1}$) in each passband. 
\item Stellar Transit Survey. 
\item Deep Survey of M31 \citep{lee12}.  
\end{enumerate}

The $w_{\rm P1}$-band Solar System Survey has contributed most to the discoveries of Solar System minor bodies due to its optimized cadence for searching moving objects and deeper limiting magnitude of 22.5, which is about one magnitude more than the $3\pi$ survey. To demonstrate the sky coverage of PS1, we separate the entire sky into 360 (R.A.) x 180 (Dec.) pixels, and each pixel is 1 square degree. Then we register the location of each pointing from the 3pi and Solar System Survey and assign it to a pixel location. We filled the eight pixels surrounding the center pixel, but then scale by a factor of 7/9, because of the 7 square degree Field-of-View of PS1.

\begin{figure}
\includegraphics[width = 1\textwidth]{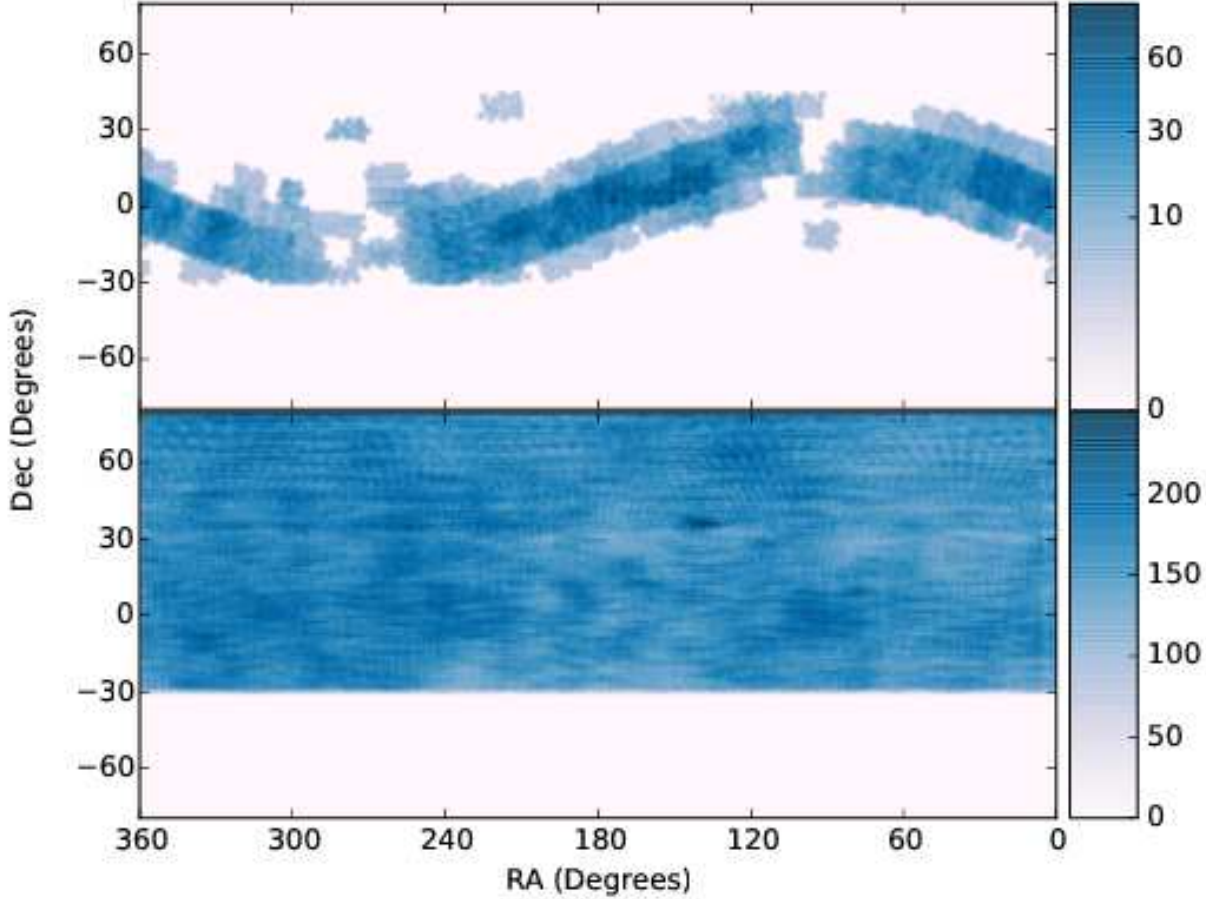}
\caption{The sky coverage of the Solar System (top) and  $3\pi$ surveys (bottom). The color bar shows the total number of exposures in all bands with the same pointing. \label{fig1}}
\end{figure}

The approximate sky coverage of the $3\pi$ and solar system survey from 2010 to 2014 is illustrated in Figure~\ref{fig1}. 
The concentration of the PS1 solar system survey within $\pm 10^{\circ}$ and part of the area between $\pm 10^{\circ}$ and $\pm 15^{\circ}$ of the ecliptic plane are clearly shown. The PS1 data products will be released to the public in 2017.

The PS1 Outer Solar System Pipeline will be described fully in forthcoming papers by Holman et. al. (2016, in prep) and Payne \& Holman (2016, in prep.). We used the PS1 Outer Solar System Pipeline to process the data and search for slow-moving Solar System objects. The PS1 Outer Solar System Pipeline uses a distance-based approach for identifying and linking point source detections. It begins with the source detection catalogs of direct images produced by the PS1 Image Processing Pipeline (IPP), rather than difference images.  We avoid using difference-image source catalogs because  slow moving objects are either eliminated, or their SNRs are significantly reduced, due to the short time interval (15 min) between two consecutive exposures. In the next step, the pipeline develops a catalog of stationary objects for elimination in each exposure.

After the removal of the stationary sources, the pipeline will identify  ``tracklets'', i.e., sequences of source detections in the same night that are consistent with linear motion in the constant rate. The pipeline evaluates the tracklets of moving objects by the goodness of fit, using estimated astrometric uncertainties. Finally, the pipeline links the tracklets over intervals ranging from a few nights to multiple years, allowing a full fit to be performed to characterize the orbital parameters.

\section{Identification and Confirmation of Trojan Candidates and the Discoveries}
\label{SECN:DET}
In the selection of NT candidates, objects with semi-major axes between 29.7 to 30.3 AU and e $<$ 0.3 were chosen from the outer Solar System pipeline. They must have at least four ``tracklets'', and two of the four must have three detections or more. Therefore, at least ten detections spread over four different nights, and the total observational arc-lengths must be longer than one year will reach our minimal criteria. For example, the rediscovery of known NT, 2011 QR$_{322}$, was just passing our minimal criteria; it has a pair detections in Oct, 2012, two triplet detections separated in different nights of Oct., 2013, and has the other pair of detections in Nov., 2013.

Because of the long observational arc-lengths, all candidates have fairly well determined orbital elements. One thousand clones of each candidate were generated from the orbit fitting covariance matrix generated by the {\it orbfit} code of \citet{ber00}. From the 1,000 clones we select the following three to be numerically integrated for 10Myr:  the best-fit, the smallest semi-major axis and the largest semi-major axis. If any of the three clones exhibits dynamical coupling with Neptune with the resonant argument, $\phi_{1:1}$, $\sim 60^{\circ}$ or $300^{\circ}$, they will be classified as a candidate NT. Note that the resonance argument, $\phi_{1:1} = \lambda_N - \lambda_T$, is defined by the difference of the mean longitude of Neptune ($\lambda_N$) and that of the Trojan candidate ($\lambda_T$) with  $\lambda = M + \Omega + \omega$, where $M$ is the mean anomaly, $\Omega$ is the longitude of the ascending node, and $\omega$ is the argument of perihelion.

Once the candidates have been identified, we checked whether they have been detected by the Dark Energy Survey \citep{dar16} or not. We also used the CADC SSOIS system \citep{gwy12} to examine whether the candidates have been observed in other archival data. One of the NTs, 2011 SO$_{277}$, was observed by the Dark Energy Survey, and another one, 2010 TT$_{191}$, was observed by CFHT in 2007. We also carried out follow-up observations at the predicted locations to confirm their existence using the Fred Lawrence Whipple Observatory 1.2m, Lulin Observatory 1m and Lijiang 2.4m telescopes of the Yunnan Astronomical Observatory. We used the astrometric data from these confirming observations to improve the orbital solutions. The same numerical procedure described above for the Trojan candidate identification was repeated. 

\begin{figure}
\includegraphics[width = 0.7\textwidth, angle =270]{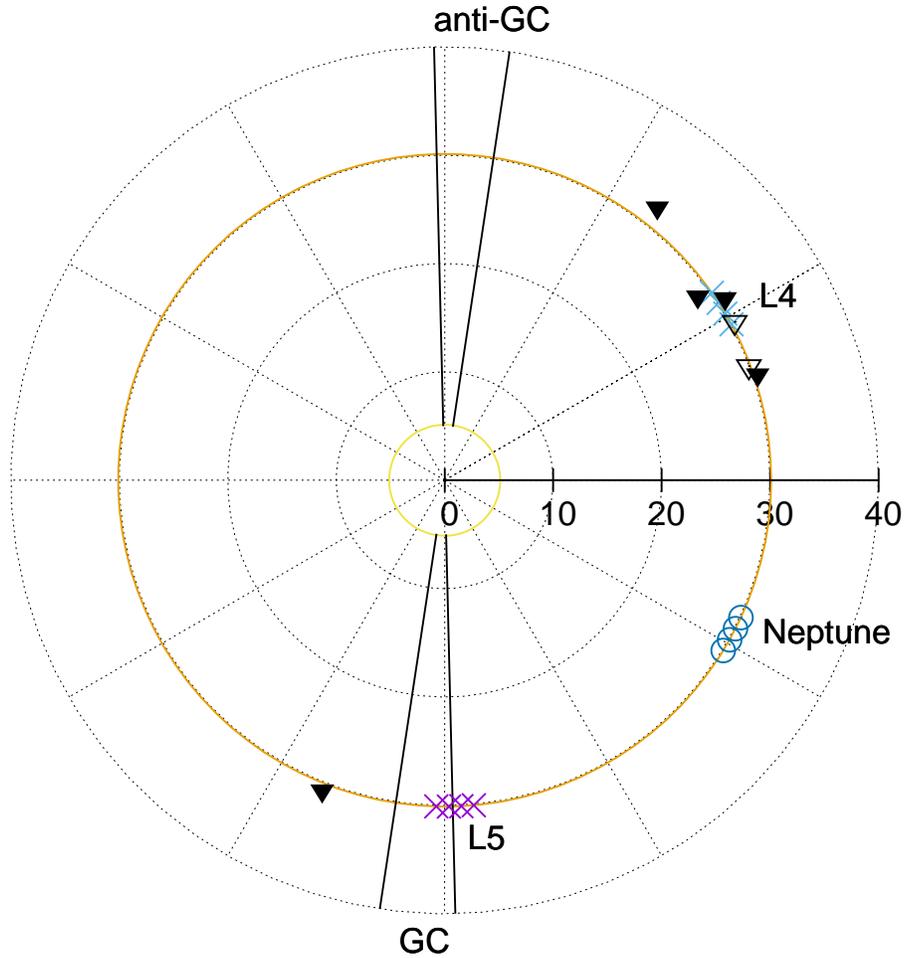}
\caption{The spatial distribution of all PS1 detected Trojans. The solid triangles are the newly discoveried NTs, and open triangles are the known ones detected by PS1. The positions of NTs correspond to their first detections of PS1. The blue circles show the locations of Neptune from 2010 to 2013, and the crosses show the corresponding Lagrange points. Notice that the Galactic Center (GC) overlapped with L5 during 2010 to 2012.  \label{fig2}}
\end{figure}

We identified seven NTs, with most of the detections being contributed by the PS1 Solar System Survey. Figure~\ref{fig2} shows their spatial distribution, and the corresponding discovery/rediscovery latitude and longitude in ecliptic coordinates can be found in Table~\ref{tab1}. The L5 region overlapped significantly with the Galactic center during the related observations. Two of the seven NTs are known L4 NTs, 2001 QR$_{322}$ and 2006 RJ$_{103}$. For the other five newly discovered Trojans, one is located at L5 and the other four at L4. The detailed observation log can be found in the Minor Planet Center database.

\section{Orbital and Physical Properties}
\label{SECN:ORB}
Table~\ref{tab1} summarizes the orbital elements of the NTs detected by PS1, including the two known ones, namely, 2001 QR$_{322}$ and 2006 RJ$_{103}$. The PS1 detections along with previous observations were used to improve the orbital elements of these two objects. It is noted that the L5 Trojan, 2013 KY$_{18}$, has a relatively large eccentricity ($e \sim 0.12$), indicating the possibility of long term orbital instability.

\clearpage
\floattable

\begin{deluxetable}{lcccccccccccc}
\tabletypesize{\scriptsize}
\tablecaption{Barycentric Oscillating Orbital Elements of PS1 detected and known NTs \label{tab1}}
\tablewidth{0pt}
\tablehead{Name & a (AU) & e & {\it i} (deg) & $\Omega$ (deg) & $\omega$ (deg) & Peri. date (JD) & Epoch (JD) &  H & L & PS1 Detected & $\beta$\tablenotemark{a} & $\lambda$\tablenotemark{b}}
\startdata
2001 QR$_{322}$ & 30.233  & 0.0285 & 1.323 & 151.636 & 158.76 & 2444677 & 2452142.8 & 7.9 & L4 & yes & 21.85 &	-1.05\\ 
2004 KV$_{18}$ & 30.353\tablenotemark{c}  & 0.189 & 13.573 & 235.593 & 295.733 & 2446125.4 & 2453351.5 & 8.9 & L5 &  & &  \\ 
385571 Otrera (2004 UP$_{10}$) & 30.184  & 0.027 & 1.431 & 34.780 & 358.452 & 2457945.4 & 2454668.5 & 8.8 & L4 &  & & \\ 
385695 (2005 TO$_{74}$) & 30.137  & 0.051 & 5.253 & 169.387 & 304.750 & 2470616.8 & 2454522.5 & 8.3 & L4 & & &  \\ 
2005 TN$_{53}$ & 30.171  & 0.064 & 24.988 & 9.278 & 85.892 & 2467102.2 & 2454775.5 & 9.0 & L4 &  & & \\ 
2006 RJ$_{103}$  & 30.038 & 0.0300 & 8.163 & 120.867 & 27.26 & 2475056 & 2453626.8& 7.5 & L4 &  yes & 28.12 & -8.42\\
2007 VL$_{305}$ & 30.004  & 0.062 & 28.125 & 188.611 & 215.518 & 2456036.1 & 2454566.5 & 7.9 & L4 &  & &\\ 
2008 LC$_{18}$ & 30.090  & 0.079 & 27.489 & 88.528 & 6.845 & 2427000.1 & 2454759.5 & 8.4 & L5 &  & & \\ 
2010 TS$_{191}$ & 30.006 & 0.0457 & 6.563 & 129.600 & 299.5 & 2460637 & 2455476.9 & 7.9 & L4 & yes & 36.07	& -6.76\\
2010 TT$_{191}$   & 30.094 & 0.0701 & 4.276 & 249.295 & 7.8 &  2429839 & 2454419.0 & 7.9 & L4 & yes & 54.76 & 1.18\\
2011 HM$_{102}$ & 30.119  & 0.081 & 29.389 & 100.993 & 152.287 & 2452480.3 & 2455758.5 & 8.1 & L5 & & &  \\ 
2011 SO$_{277}$   & 30.161 & 0.0118 & 9.639 & 113.528 & 117.7 & 2431675  & 2455831.0 & 7.6 & L4 &  yes & 16.19 & -9.87\\ 
2011 WG$_{157}$   & 30.031 & 0.0278  & 23.299 & 352.165 & 215.3 & 2482896 & 2455885.8 & 7.0 & L4 & yes &  41.95 & 18.06 \\
2012 UV$_{177}$ & 30.175  & 0.074 & 20.811 & 265.753 & 200.784 & 2467673.8 & 2456131.5 & 9.2 & L4 & & & \\ 
2013 KY$_{18}$    & 30.149 & 0.123 & 6.659 & 84.397 & 271.2 & 2471956 & 2456429.0 & 6.6 & L5 & yes & 249.82 & 1.79 \\
2014 QO$_{441}$ & 30.104  & 0.105 & 18.824 & 107.110 & 113.897 & 2429010.4 & 2456910.5 & 8.3 & L4 &  & &\\ 
2014 QP$_{441}$ & 30.0785  & 0.067 & 19.394 & 96.626 & 2.639 & 2467286.1 & 2456979.5 & 9.3 & L4 &  & &\\ 
\enddata
\tablenotetext{a}{The discovery ecliptic Latitude}
\tablenotetext{b}{The discovery ecliptic Longitude}
\tablenotetext{c}{The barycentric orbital elements of Non-PS1 detected NTs were queried from JPL HORIZONS System. }
\end{deluxetable}

To understand the orbital properties and resonant behaviors of these Trojan candidates, we produced 1,000 clones for each Trojan candidate covering the error ellipse of its orbital elements. That is, the initial orbital elements of each clone were generated from a multivariate normal distribution with the six dimensional covariance matrix provided by the observation fitting routine, i.e., the {\it Orbfit} code of \citet{ber00}. Forward integration was performed for each clone over a time interval of 1 Gyr using the {\tt Mercury 6.2} N-body code of \citet{cha99}. 
 As heliocentric orbital elements are used as the standard input to {\tt Mercury 6.2}, the {\it orbfit} code has been modified to generate heliocentric orbital elements and the corresponding covariance matrix.

Table~\ref{tab2} shows the mean orbital elements, half-peak RMS libration amplitudes, libration periods and lifetimes of the seven NTs. These orbital parameters were computed from the numerical results of the first 100 Myr of the orbital integration except for 2013 KY$_{18}$ where the numerical values were derived from the first million years due to its short dynamical life time.

\begin{figure}
\includegraphics[width = 0.7\textwidth, angle = 270]{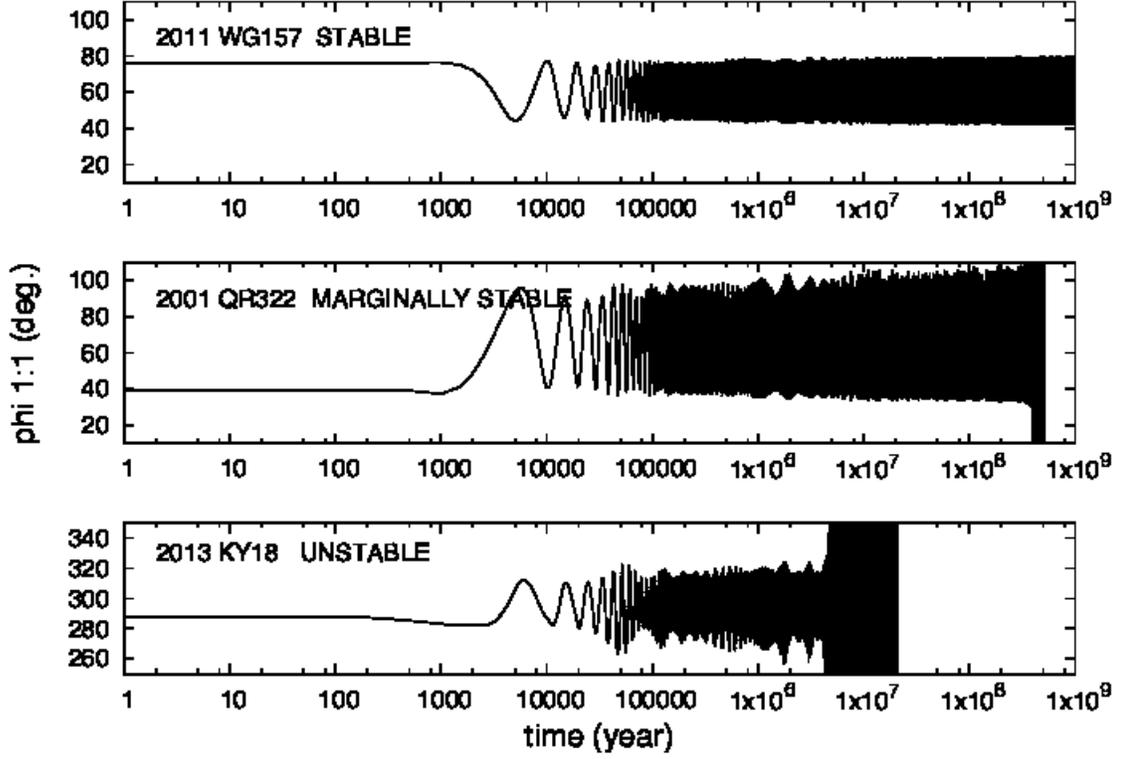}
\caption{The resonant argument variations of 2011 WG$_{157}$ (top), 2001 QR322 (middle) and 2013 KY$_{18}$ (bottom). These represent three types of orbital stability, namely Stable, Marginally Stable and Unstable Neptune Trojans respectively. \label{fig3}}
\end{figure}

Our calculations show that most of the L4 NTs, except 2001 QR$_{322}$, have a half-life longer than 1 Gyr.  For example, all of the clones of 2011 WG$_{157}$, 2010 TS$_{191}$ and 2006 RJ$_{103}$, remained stable during the entire 1 Gyr orbital integration. Conversely, 2011 SO$_{277}$ and 2010 TT$_{191}$ lost about 100 (10$\%$) and 300 (30$\%$) clones, respectively. The known NT, 2001 QR$_{322}$ was found to have a half-life of about 0.53 Gyr which agrees well with the previous results of about 0.55 Gyr from \citet{hor10a}. The only L5 Trojan, 2013 KY$_{18}$, has a short half-life of about 3.2 Myr which is similar to the value of less than 1 Myr of another unstable L5 Trojan, 2004 KV$_{18}$ \citep{gua12, hor12b}. This suggests that 2013 KY$_{18}$ is likely to be a temporarily captured Trojan.
Figure~\ref{fig3} shows the variations in resonant argument, $\phi_{1:1}$, of 2011 WG$_{157}$ (top), 2001 QR$_{322}$ (middle) and 2013 KY$_{18}$ (bottom), which represent stable, marginally stable and unstable NTs respectively. 


\floattable
\begin{deluxetable}{lcccccc}
\tabletypesize{\scriptsize}
\tablecaption{Barycentric Orbital Properties of PS1 detected Neptune Trojans  \label{tab2}}
\tablewidth{0pt}
\setlength{\tabcolsep}{0.08in} 
\tablehead{Name & $<$a$>$\tablenotemark{a} (AU) & $<$e$>$ & $<$i$>$ (deg) & Libration ampl. (deg) & Libration per. (year) & Half-life time }
\startdata
2001 QR$_{322}$ & 30.107 $\pm$\tablenotemark{b} 0.122 & 0.030 $\pm$ 0.009 & 1.90 $\pm$ 0.82 & 27.3 $\pm$\tablenotemark{c} 0.1 & 9268 $\pm$ 0 & 0.53 Gyr \\ 
2006 RJ$_{103}$  & 30.106 $\pm$ 0.027 & 0.025 $\pm$ 0.008 & 6.76 $\pm$ 1.12 & 5.6 $\pm$ 0.2 & 8858 $\pm$ 3 &  $>$ 1 Gyr \\
2010 TS$_{191}$   & 30.106 $\pm$ 0.056 & 0.047 $\pm$ 0.008 & 5.11  $\pm$ 1.11 & 12.1 $\pm$ 0.6 & 8896 $\pm$ 11 &  $>$ 1 Gyr \\
2010 TT$_{191}$   & 30.107 $\pm$ 0.091 & 0.068 $\pm$ 0.008 & 5.93  $\pm$ 1.10 & 19.9 $\pm$ 1.2 & 9040 $\pm$ 30 &  $>$ 1 Gyr \\
2011 SO$_{277}$   & 30.107 $\pm$ 0.089 & 0.016 $\pm$ 0.006 & 7.84  $\pm$ 1.16 & 19.7 $\pm$ 1.9 & 9104 $\pm$ 51 &  $>$ 1 Gyr \\
2011 WG$_{157}$  & 30.106 $\pm$ 0.068 & 0.027 $\pm$ 0.009 & 23.11 $\pm$ 1.12 & 15.6 $\pm$ 0.1 & 9458 $\pm$ 4 &  $>$ 1 Gyr \\
2013 KY$_{18}$ & 30.107 $\pm$ 0.095 & 0.106 $\pm$ 0.008 & 4.70  $\pm$ 0.98 & 20.8 $\pm$ 1.6 & 9023 $\pm$ 44 & 3.2 Myr \\
\enddata
\tablenotetext{a}{Mean elements, libration amplitude and  libration period were calculated from the results of forward 10Myrs integrations, except 2013 KY$_{18}$, was calculated from the results of forward 1Myrs integrations.}
\tablenotetext{b}{$\pm$ of mean {\it a, e, i} are the half-peak RMS.}
\tablenotetext{c}{$\pm$ of Libration amplitude and Libration period were calculated from the standard deviations from 1000 clones.}

\end{deluxetable}

\newpage

\newpage

\section{Inclination Distribution of stable L4 NTs}

\begin{figure}
\includegraphics[width = 1\textwidth]{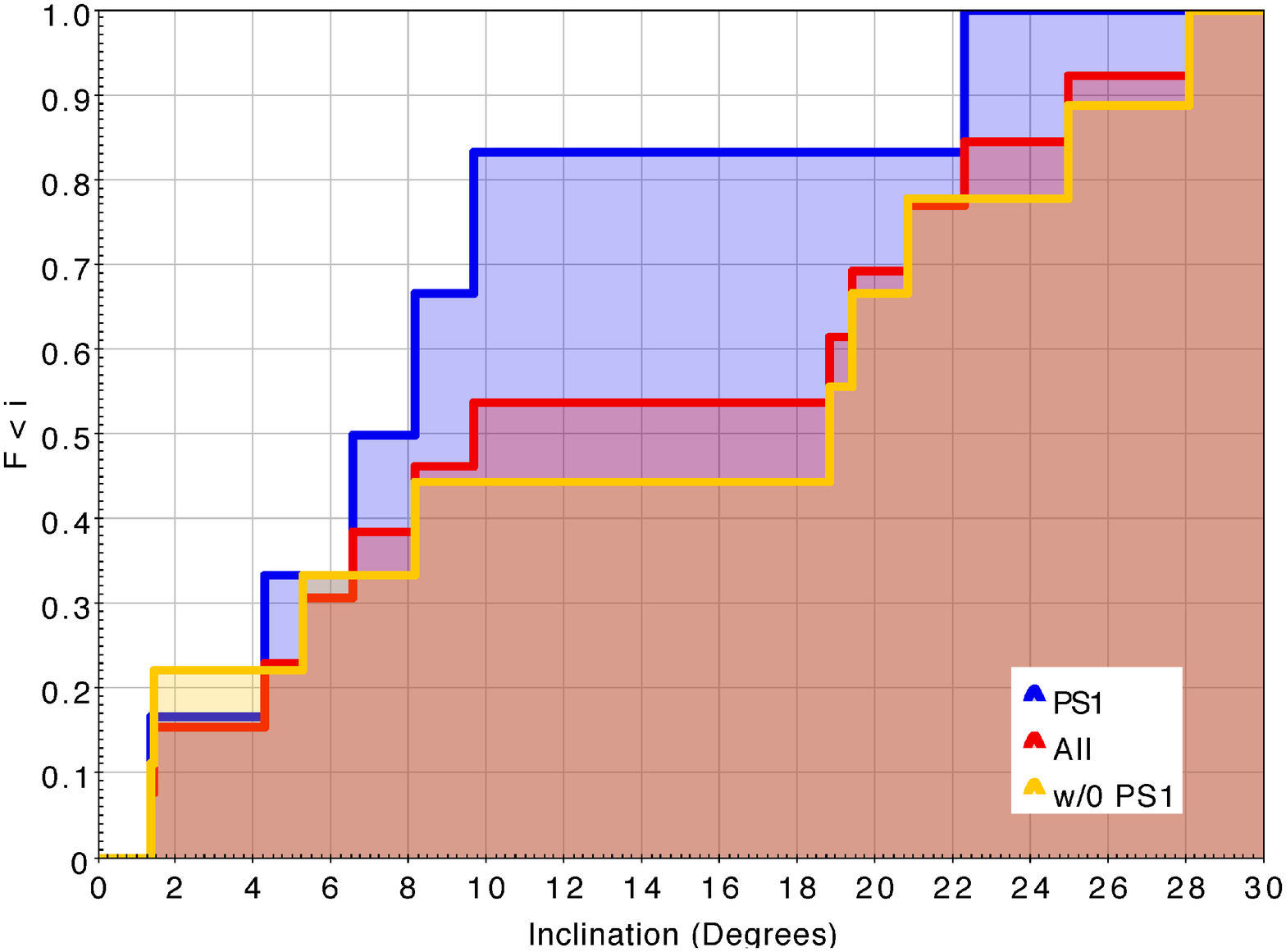}
\caption{The cumulative inclination distributions of Neptune Trojans for PS1 detected only (Blue), all known Trojans included (Red) and excluded PS1 discoveries (Yellow). We find that (a) there are no Neptune Trojans with $10^{\circ}<i<18^{\circ}$, and (b) the PS1 Neptune Trojans have a low inclination distribution.  \label{fig4}}
\end{figure}

From the dynamical stability test described in section \ref{SECN:ORB}, six of the seven NTs have Half-life time longer than 0.5 Gyr. That is, they belong to the stable population. Figure~\ref{fig4} shows the cumulative inclination distribution of those six objects. Two things are noticeable: (1) the presence of a bi-modal inclination distribution without stable L4 TNs with inclinations between 10 to 18 degree. (2) The NTs detected by PS1 display a rather low inclination distribution.

\citet{par15} first suggested that the inclination distribution of Neptune Trojans might exhibit a bimodal structure from the observational point of view. This bimodal inclination distribution may have a dynamical origin. \citet{zho09} demonstrated that the equality between the frequencies of $f_{2N:1U} - 2f_{\sigma}$ and the fundamental secular frequency $g_6$ can cause an instability, and its effective region crosses 15 degrees in inclination when a $\sim$ 30.1 AU.  Here $f_{2N:1U}$ is the frequency of the quasi-2:1 mean motion resonance between Neptune and Uranus, and $f_{\sigma}$ is the libration frequency. This dynamical effect could explain our observations of a lack of stable Trojans between {\it i} = $10{^\circ}$ and $18^{\circ}$. 

In our PS1 survey, we have detected only one stable L4 Trojan with inclination greater than $20^{\circ}$ and observed a rather low inclination distribution of NTs. To compare our observational result with the intrinsic inclination distribution estimated in \citet{par15}, we would have to debias the PS1 survey data. However, the complicated PS1 survey cadence and camera structure make a detailed study difficult at the present moment. In the following, a simplified procedure is used to estimate the survey bias in inclination space. 

The point of the inclination debiasing is to estimate what fraction of objects can be found in the PS1 survey, with our search algorithm, for a given orbital inclination. While the limiting magnitude and detection efficiency function are sufficient, especially because the NT's orbits are approximately circular, the sky coverage and number of exposures are the key factors. Therefore, our approximate PS1 detection efficiency function was assumed to be a function of the number of total exposures in a given survey region; survey regions with more exposures will have higher detection rates. Thus, the PS1 survey has the highest detection rate within $\pm 10^{\circ}$ of the ecliptic plane and a lower detection rate in the region between $10^{\circ}$ to $15^{\circ}$  and $-10^{\circ}$ to $-15^{\circ}$ above and below, respectively. To proceed, we first assume that the limiting magnitude is 22.5, and indeed all of the Neptune Trojans were detected around the limiting magnitude. Second, the detectability of m=22.5 is assumed to be 50$\%$, and the filling factor is 70$\%$. For a NT, we will therefore have a 35$\%$ chance to detect it. Third, we must detect a NT for at least 10 times before claiming that we have found it. This assumption is based on the simplify version of our detection criteria in section \ref{SECN:DET}. Hence, The PS1 detection efficiency function can be approximated as the total sum of probability mass functions of binomial distribution:
\begin{equation}
f_{eff}(n) = \sum\limits_{i=10}^n\binom{n}{i}0.35^i\times(1-0.35)^{n-i}
\end{equation}

Here, $n$ is the total number of exposures in a specific survey region, and $i$ is the minimal number of detections required for finding an object in our detecting pipeline. 
The result will be, for example, if a survey region has 20, 30 and 40 exposures, the detection efficiency will be $\sim$ 0.12, 0.64 and 0.94, respectively.
Using this detection efficiency function along with the approximated sky coverage map of the PS1 Solar System Survey (see Figure~\ref{fig1}), we would be able to compute the whole sky detectability of NTs in 1 square degree resolution.

Having estimated the PS1 detection efficiency function, we can use the survey simulator from the Outer Solar System Origins Survey (OSSOS) \citep{ban15, kav09} in combination with the NT population model given in \citet{par15}, in which the intrinsic inclination ($i$) distribution is equivalent to a truncated Brown's distribution,
\begin{equation}
p(i) = \left\{
\begin{array}{lcl}
\sin(i)\exp(-\frac{1}{2}(i/\sigma_{i})^2)di, & & {i < i_t} \\
0, & & {i \geq i_t},
\end{array}
\right.
\end{equation}
, the intrinsic libration amplitude ($L_{11}$) distribution is a truncated Rayleigh distribution,
\begin{equation}
p(i) = \left\{
\begin{array}{lcl}
L_{11}\exp(-\frac{1}{2}(L_{11}/\sigma_{L_{11}})^2)dL_{11}, & & {L_{11} < L_{11t}} \\
0, & & {L_{11} \geq L_{11t}},
\end{array}
\right.
\end{equation}
and finally the intrinsic eccentricity distribution also follows a truncated Rayleigh distribution,
\begin{equation}
p(i) = \left\{
\begin{array}{lcl}
e\exp(-\frac{1}{2})(e/\sigma_{e})^2)de, & & {e < e_{t}} \\
0, & & {e \geq e_{t}}.
\end{array}
\right.
\end{equation}
Here $i_{t}$, $L_{11t}$ and $e_{t}$ are the truncation points of the inclination, libration amplitude and eccentricity distributions, respectively. 

\begin{figure}
\includegraphics[width = 1\textwidth]{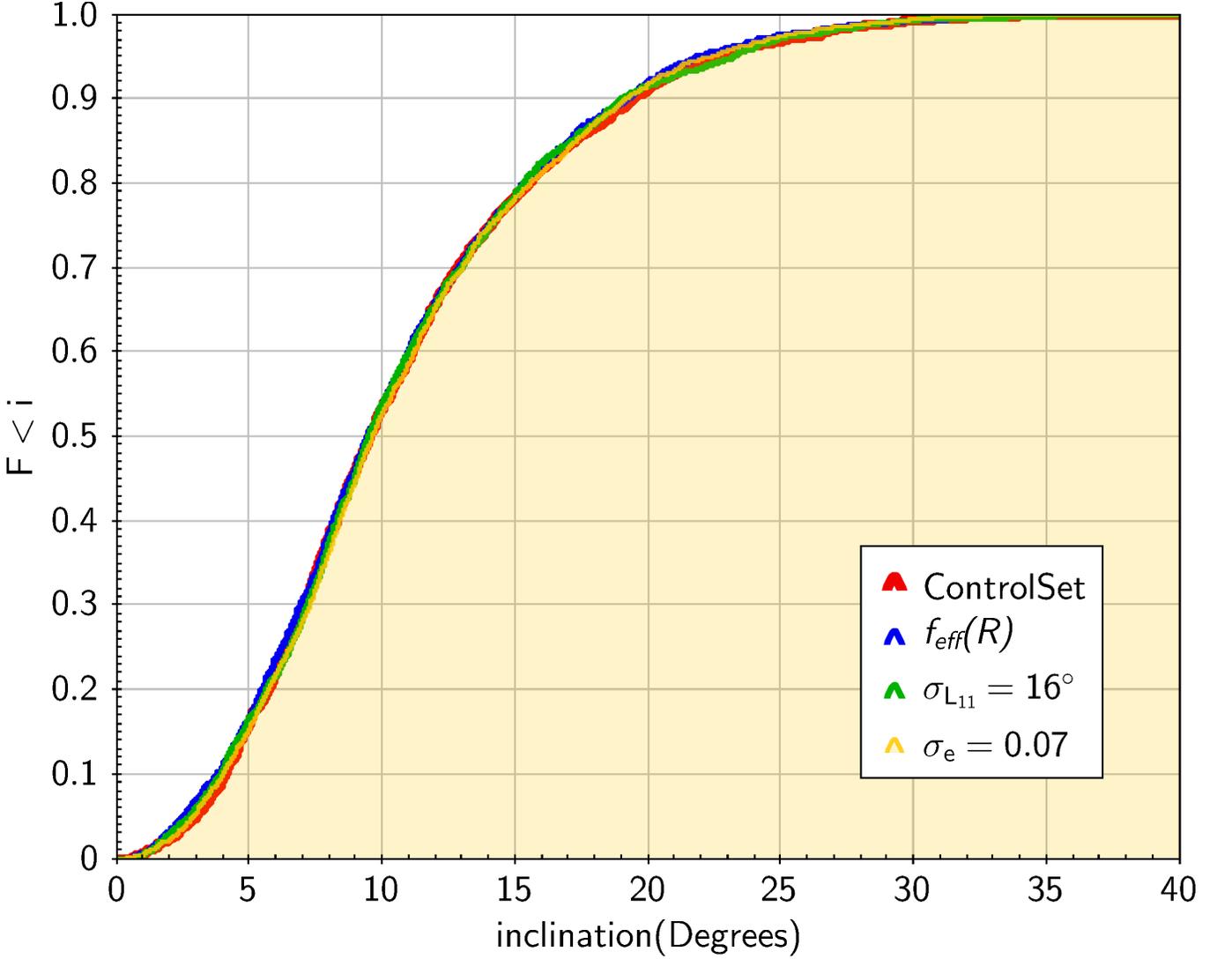}
\caption{The simulated biased inclination distributions with different detection efficiency function or NT population model parameters. The results show that the simulated biased inclination distribution is independent with these factors. \label{fig5}}
\end{figure}

To investigate the correlation between $\sigma_{i}$ and the detection efficiency function during the survey simulations, we first ran the simulator with two different detection efficiency function, (1) our brightness independent PS1 detection efficiency function, and (2) a double hyperbolic tangents brightness dependent detection efficiency function \citep{pet06}: 
\begin{equation}
f_{eff}(R) = \frac{A}{4}[1-tanh(\frac{R-R_c}{\Delta_1})][1-tanh(\frac{R-R_c}{\Delta_2})]
\end{equation}
Here, A, R$_c$, $\Delta_1$ and $\Delta_2$ are the filling factor (or maximal efficiency), roll-over magnitude ($50\%$ of the maximal efficiency), and widths of the two components, respectively. In this simulation, we set A = 0.9, R$_c$ = 22.5, $\Delta_1$ and $\Delta_2$ are 0.01 and 0.15, respectively. We also made the similar procedure but with different $\sigma_{L_{11}}$ and $\sigma_{e}$. The parameters of each simulation are shown in table~\ref{tab3}, and the results of the simulated biased inclination distributions were shown in Figure~\ref{fig5}. 

\floattable
\begin{deluxetable}{lcccccc}
\tabletypesize{\scriptsize}
\tablecaption{Parameters of Survey Simulations  \label{tab3}}
\tablewidth{0pt}
\setlength{\tabcolsep}{0.08in} 
\tablehead{Simulation ID &$\sigma_{i} (^{\circ})$ & $\sigma_{L_{11}} (^{\circ})$ & $\sigma_{L_{11t}} (^{\circ})$ & $\sigma_{e}$ & $\sigma_{e_t}$ & Efficiency function}
\startdata
Control Set & 11 & 10 & 35 & 0.044 & 0.12 & $f_{eff}(n)$\tablenotemark{a} \\
$f_{eff}(R)$ & 11 & 10 & 35 & 0.044 & 0.12 & $f_{eff}(R)$\tablenotemark{b} \\
$\sigma_{L} = 16^{\circ}$ & 11 & 16 & 35 & 0.044 & 0.12 & $f_{eff}(n)$ \\
$\sigma_{e} = 0.07$ & 11 & 10 & 35 & 0.07 & 0.12 & $f_{eff}(n)$ \\
\enddata
\tablenotetext{a}{brightness independent Detection efficiency function (Equation 4)}
\tablenotetext{b}{brightness dependent Detection efficiency function (Equation 8)}
\end{deluxetable}

The results clearly shows that changing $\sigma_{L_{11}}$, $\sigma_{e}$ or using different detection efficiency function do not affect the simulated biased inclination distributions.
Therefore, to test the intrinsic inclination distribution, we set the $\sigma_{i}$ from $5^{\circ}$ to $21^{\circ}$ with the 1 degree steps, and fixed the $\sigma_{L_{11}}$, $\sigma_{L_{11t}}$, $\sigma_{e}$ and $\sigma_{e_t}$, respectively, to be $10^{\circ}$, $35^{\circ}$, 0.044 and 0.12 as suggested by \citet{par15} in our simulations.

\begin{figure}
\includegraphics[width = 0.7\textwidth, angle = 270]{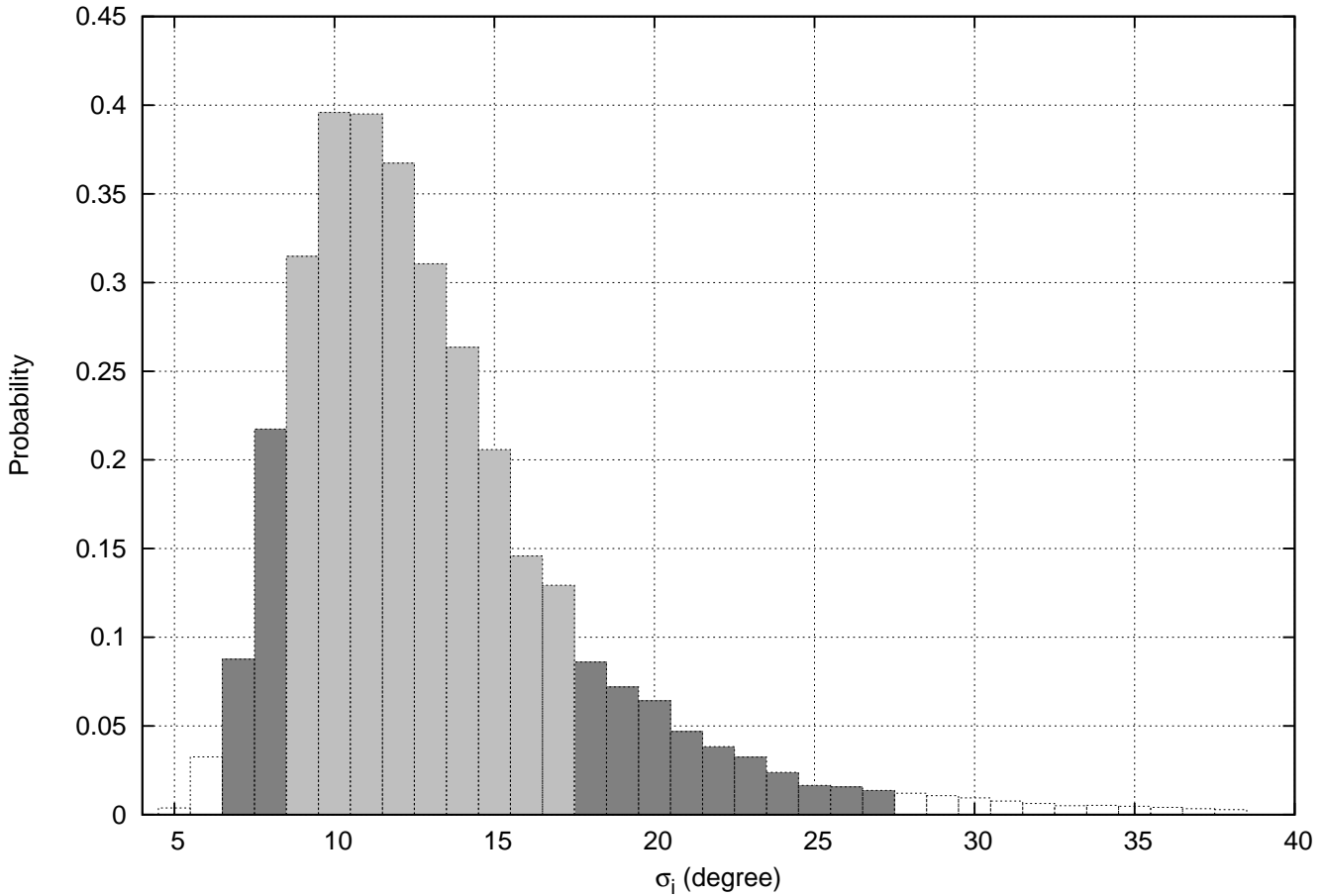}
\caption{The probability distribution of Neptune Trojan inclination width parameter $\sigma_{i}$. Here the probability is the chance to reproduce  high-{\it i}/low-{\it i} = 1/5 with specific $\sigma_i$. The shaded areas illustrate the $68\%$ ( $9^{\circ} \leq \sigma_{i} \leq 17^{\circ}$) and $95\%$ ( $7^{\circ} \leq \sigma_{i} \leq 27^{\circ}$) confidence intervals determined directly from the distribution. \label{fig6}}
\end{figure}

To compare the simulation results with our observation, we computed the ratio of the high-{\it i} ($i > 18^{\circ}$) to low-{\it i} ratio ($i < 10^{\circ}$) NTs from the simulation results, and estimated the probability to generate the high-{\it i}/low-{\it i} ratio equal to 1/5 that was obtained from the PS1 survey. The criterion of the low- and high-{\it i} follows the from fact that currently there are no known NT within the 10 to 18 degree inclination range. The results are shown in Figure~\ref{fig6}. The shaded areas show the  $68\%$ confidence (1 sigma) and and $95\%$ confidence (2 sigma) intervals. The cases of $\sigma_{i} > 27^{\circ}$ and $\sigma_{i} < 7^{\circ}$ would be rejected at 2 sigma level, and the most likely value is $\sigma_{i} \sim 11^{\circ}$. 

\section{Luminosity function of stable L4 NTs}
Our approximate PS1 detection efficiency is a function of total number of exposures in a specific survey region. For the NTs we found in the PS1 survey, the detection probabilities can be estimated from their detected coordinates. The detection probabilities of six L4 NTs are listed in Table~\ref{tab4}. 

\floattable
\begin{deluxetable}{lccc}
\tabletypesize{\scriptsize}
\tablecaption{Detection probabilities of six L4 NTs \label{tab4}}
\tablewidth{0pt}
\setlength{\tabcolsep}{0.08in} 
\tablehead{Name & H & prob. & 1/prob.}
\startdata
2001 QR$_{322}$ & 7.9 & 1 & 1 \\ 
2006 RJ$_{103}$ & 7.5 & 1 & 1\\
2010 TS$_{191}$ & 7.9 & 0.45 & 2.2\\
2010 TT$_{191}$ & 7.9 & 0.365 & 2.74\\
2011 SO$_{277}$ & 7.6 & 0.8 & 1.25\\
2011 WG$_{157}$ & 7.0 & 1 & 1\\
\enddata
\end{deluxetable}

We divided the L4 NT population into two bins, H $< 7.5$ and H $>7.5$, and the density in each bin is the sum of the 1/prob. of every objects in that bin, which is 2 and 7.18, respectively. The approximate slope, $\alpha$, is log$_{10}$((7.18/2) /$\Delta$H)) $\sim 0.86$, where $\Delta$H is the H difference between two bins.
This results shows that our rough debiasing produces an slope consistent with the values from \citet{she10b} and \citet{fra14}.

\section{Discussion}

\citet{par15} suggested that if the stable NT cloud follows an inclination distribution similar to that of the Jovian Trojan population, the corresponding inclination width must be greater than $11^{\circ}$. 
Our result, which is based on six stable L4 NTs, is roughly consistent with his finding. 
Note that our most likely value of $\sigma_{i} \sim 11^{\circ}$ is the minimal acceptable value in \citet{par15}. Therefore, our present result might be indicative of a lower inclination distribution.

It is worth noting that a high and wide NT inclination distribution with $\sigma_{i} \sim 20^{\circ}$ is unlikely to result from capture from a dynamically cold disk without orbital damping during planet migration. However, the scenario is possible if the actual NT inclination distribution has $\sigma_{i}$ only around $10^{\circ}$ \citep{nes09, par15, che16}.

One other fact that should be taken into consideration is that
the PS1 survey can only detect larger NTs ($H \lesssim 8$) compared to the other surveys with fainter limiting magnitudes. It may be the case that large and small NTs have different high/low-{\it i} ratios: If the NT cloud actually has cold and hot populations like the classical Kuiper Belt and the two populations have different size distributions, it might also explain the inconsistent measurements of the high/low inclination ratio.

\citet{par15} and \citet{che16} simulated the captured NTs after planet migration and found that there is no difference between the numbers of captured L4 and L5 Trojans.
The PS1 survey should not have any bias to detect high eccentricity objects in L4 the region. However, we did not detect any unstable L4 NTs with high eccentricity. 
In the near future, as the L5 region moves away from the Galactic center, we will be able to test the possible asymmetry between the L4 and L5 populations. The ongoing PS1 + PS2 survey would be able to cover more than $\pm 20^{\circ}$ above and below the ecliptic plane, and will be very useful in deriving a less-biased inclination distribution of NTs. In addition, the future LSST survey will detect many more NTs, allowing a more nuanced understanding of their distribution to be gained.


\section{Summary}

We report the detection of seven Neptune Trojans in the PS1 Outer Solar System Survey. Five of these are new discoveries and consist of one L5 Trojan and four L4 Trojans. 
Our numerical integrations show that the new L5 Trojan can be stable for only 3.2 Myr, and suggest that it is a temporarily captured object. The four new L4 Trojans can remain stable for over 1 Gyr and could be members of a primordial population. Only one stable L4 Trojan with inclination higher than $20^{\circ}$ was detected by the PS1 survey. Our survey simulation results show that if the L4  NT cloud follows an inclination distribution similar to that of the Jovian Trojan population,  at $>$ 95\% confidence, it should have an inclination width, $\sigma_{i}$, between $7^{\circ}$ and $27^{\circ}$.
We suggest that the most likely value of $\sigma_{i}$ is $11^{\circ}$, which corresponds to the minimal accepted value of  $\sigma_{i}$ from  \citet{par15}. 
Compared to previous surveys that discovered other known NTs, PS1 can only detect relatively large ($H \lesssim 8$) objects. Thus, the possible inconsistency between \citet{par15} and our result could be a hint of a size-dependent inclination distribution of NTs.


\acknowledgments

We would like to thank Zhong-Yi Lin and JianGuo Wang for observing with Lijiang 2.4m telescope.
We are garteful to Mike Alexandersen for providing the Neptune Trojan population model of the OSSOS survey simulator.
We are grateful to Gareth Williams of the Minor Planet Center for his ongoing help.

This work was supported in part by MOST Grant: MOST 104-2119-008-024 (TANGO II) and MOE under the Aim for Top University Program NCU, and Macau Technical Fund: 017/2014/A1 and 039/2013/A2. HWL acknowledges the support of the CAS Fellowship for Taiwan-Youth-Visiting-Scholars under the grant no. 2015TW2JB0001.

The Pan-STARRS1 Surveys (PS1) have been made possible through contributions of the Institute for Astronomy, the University of Hawaii, the Pan-STARRS Project Office, the Max-Planck Society and its participating institutes, the Max Planck Institute for Astronomy, Heidelberg and the Max Planck Institute for Extraterrestrial Physics, Garching, The Johns Hopkins University, Durham University, the University of Edinburgh, Queen's University Belfast, the Harvard-Smithsonian Center for Astrophysics, the Las Cumbres Observatory Global Telescope Network Incorporated, the National Central University of Taiwan, the Space Telescope Science Institute, the National Aeronautics and Space Administration under Grant No. NNX08AR22G issued through the Planetary Science Division of the NASA Science Mission Directorate, the National Science Foundation under Grant No. AST-1238877, and the University of Maryland.

\vspace{5mm}
\facilities{PS1, CFHT, DECam,  FLWO 1.2m, Lijiang 2.4m, LOT}
\software{orbfit \citep{ber00} and Mercury 6.2 \citep{cha99}}


\end{CJK*}
\end{document}